\documentclass[12pt,preprint]{aastex}
\usepackage{float,epsfig,psfig}

\newcommand{\GA}{\mbox{\raisebox{-0.6ex}{$\stackrel{\textstyle>}{\sim}$}}}
\newcommand{\cxo}{{\sl Chandra}}

\newcommand{\elip}{{$D_{25}$}}
\newcommand{\lfir}{$L_{\rm FIR}$}
\newcommand{\lx}{$L_{\rm X}$}
\newcommand{\lb}{$L_{\rm B}$}

\begin{document}

\title{
A Study of Ultra-Luminous X-ray Sources from the
Chandra Archive of Galaxies\\
}

\author{
Douglas~A.~Swartz\altaffilmark{1},
Kajal~K.~Ghosh\altaffilmark{1},
Allyn~F.~Tennant\altaffilmark{2} \\ 
{\it Presented to the American Astronomical Society Meeting 201, 54.13}
}
\altaffiltext{1}{Universities Space Research Association,
NASA Marshall Space Flight Center, SD50, Huntsville, AL, USA}
\altaffiltext{2}{Space Science Department,
NASA Marshall Space Flight Center, SD50, Huntsville, AL, USA}


\begin{abstract}

The more than 80 nearby galaxies imaged with the 
Chandra Advanced CCD Imaging Spectrometer have been analyzed in a search for 
Ultra-Luminous X-ray (ULX) sources. The sample of galaxies span the range of 
Hubble morphological types and include galaxies of various mass, gas content, 
and dynamical state. X-ray characteristics of the resulting ensemble of ULX 
candidates are reported and correlations with properties of the host galaxies 
are presented.

Support for this research was provided in part by 
NASA/Chandra grant AR2-3008X to D.A.S.

\end{abstract}

\section{Introduction}

Among the most intriguing objects in the X-ray sky are the non-nuclear 
Ultra-Luminous X-ray (ULX) sources in nearby galaxies. This name describes 
those sources considerably more luminous than expected for a 
spherically-accreting  object of typical neutron star mass. Here, we 
define ULX sources to be those with apparent (i.e., assumed isotropically 
emitting) intrinsic luminosities in excess of $10^{39}$ erg/s in the 0.5-8.0 keV
bandpass.

Through the first 2 years or so of operation, the Chandra X-ray Observatory (CXO) 
has imaged enough nearby galaxies using the ACIS CCD Imaging Spectrometer to 
undertake a systematic and uniform analysis of their ULX population. 
Ultimately, we wish to know the full pedigree of these extreme objects: 
what are their origins and history; why and how do they differ from their 
more-common low-lumiosity cousins; what does the population of ULX sources 
reveal about the nature of galaxy formation and evolution; and what influence 
do ULX sources have on their local environments? Here, we report principally 
the correlations between ULXs and global properties of their host galaxies. 

\section{X-Ray Data Reduction Methods}

Based on integration time and the best distance estimate available, 
we selected all those galaxies for which $\sim$100 source counts are expected 
from a ULX. In integration time units of ks and distances in Mpc, this 
corresponds to observations with $t/D$\GA 0.12.  
To date, 85 galaxies in the CXC public archive meet this selection criterion.

For all candidate galaxy ACIS images, the following steps were taken:
\begin{itemize}
\item All events within the \elip\  isophote were extracted 
from Level 2 event files. 
\item X-ray sources (to $\sim$3.5 S/N) were located using standard methods. 
\item Source and local background spectra in the 0.5~--~8.0 keV energy range were 
extracted.
\item Source count rates were determined and the source list ordered by 
decreasing count rate.
\item The (binned) time series for each source and for the entire image 
field were constructed (the latter to help identify and eliminate high noise 
level intervals).
\item Beginning with highest count-rate sources, simple models were fit to 
the spectra to establish spectral shapes and source luminosities. 
This process was extended to sources well below the 
$10^{39}$ erg/s ULX lower limit to ensure 
completeness of the sample for each galaxy. Sources within 5\arcsec\ of the host 
galaxy center were omitted from consideration.
\item Source positions were overlaid on optical images (usually DSS) to 
crudely map source locations to morphological features of host galaxies 
and to help eliminate obvious foreground objects. Similarly, ULX candidate 
source positions were queried using the NASA/IPAC Extragalactic Database 
to help eliminate known background QSOs.
\end{itemize}

\section{The Sample of Galaxies}

The CXO sample of galaxies is compared to the Tully catalogue of 2368 nearby 
galaxies in the two figures shown above. The CXO sample is composed of mostly 
nearby galaxies (Figure~1, left, note the large step at the Virgo cluster distance), 
with 1/2 the sample within 10 Mpc, and they are typically brighter in Blue 
luminosity (Figure~1, right). Throughout this work, the blue luminosity is used as a 
proxy for galaxy mass. Although blue light is more sensitive to the properties 
of the stellar population of the host galaxy than is visible light, 
the B-V colors of the CXO sample all fall within the range of  0.5 to 1.0 
magnitudes and so $B$ is a good measure of mass.

\begin{figure*}[h]
\begin{center}
\hspace{30pt}
\includegraphics[angle=-90,width=2.7in]{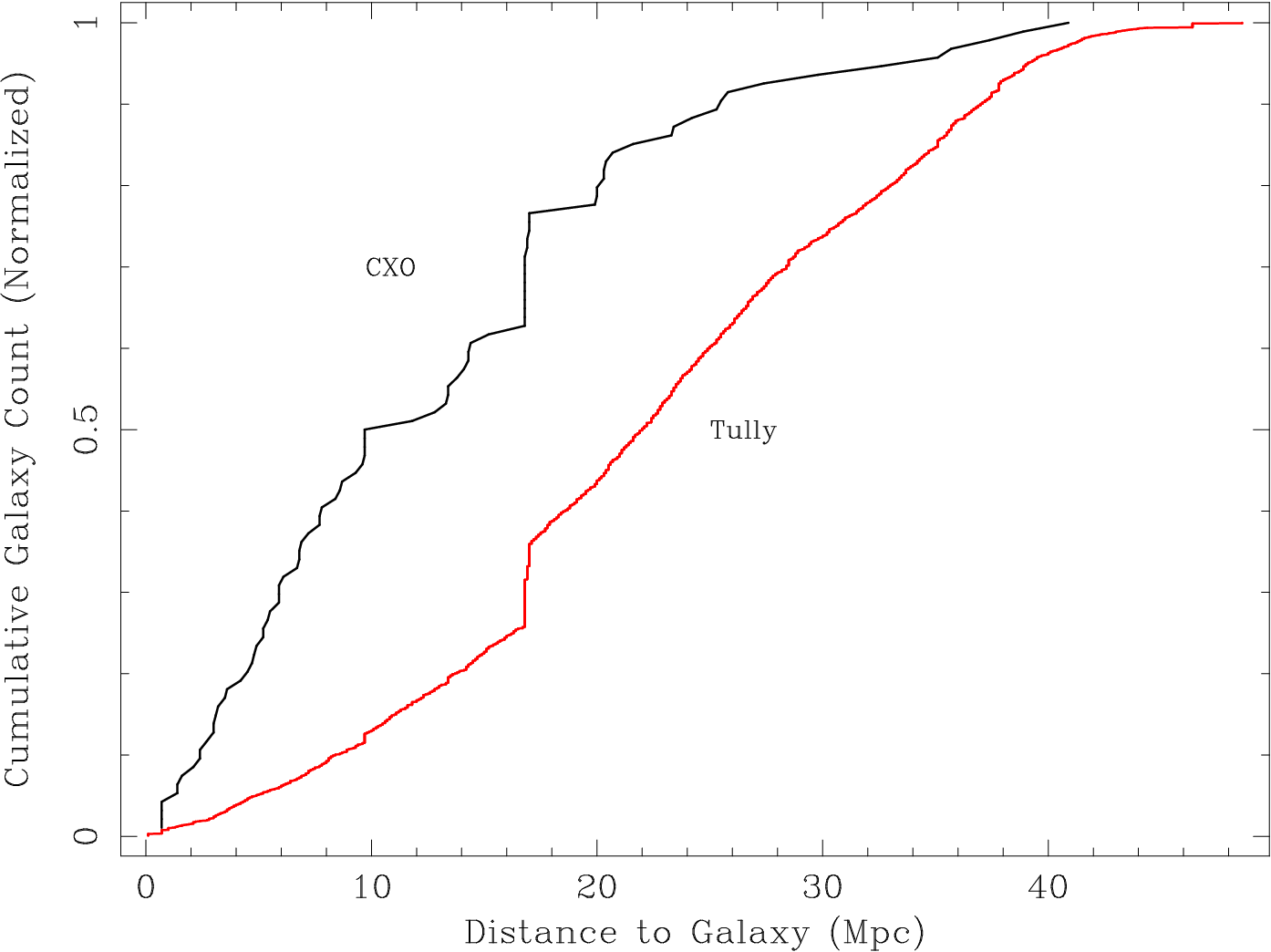}
\hspace{30pt}
\includegraphics[angle=-90,width=2.7in]{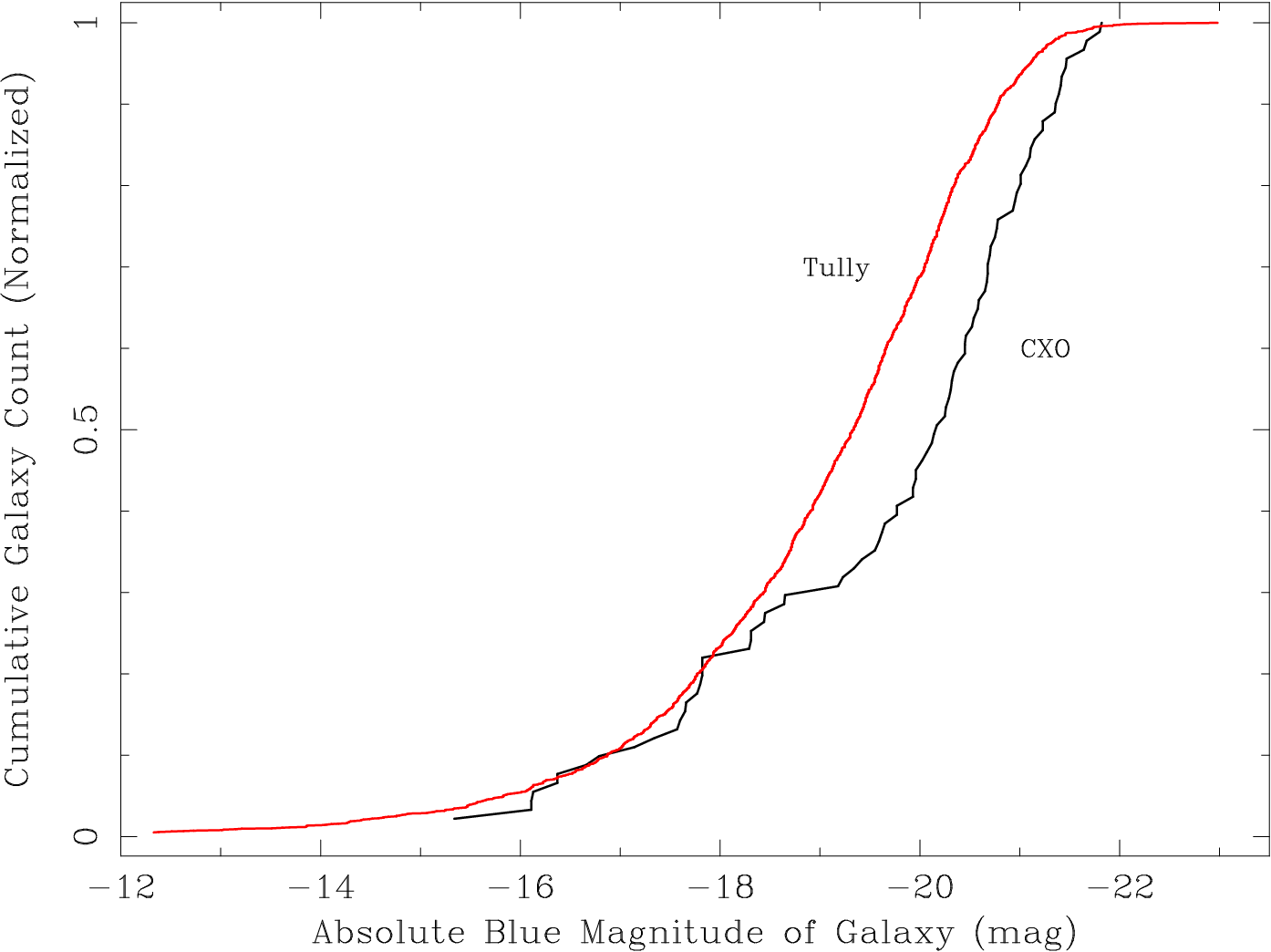}
\figcaption{Cumulative distributions of the \cxo\ sample of galaxies compared 
to the Tully (1988) catalogue over distance
({\em left}) and blue luminosity ({\em right}).
}
\end{center}
\end{figure*}
%
\begin{center}
\includegraphics[angle=-90,width=4.5in]{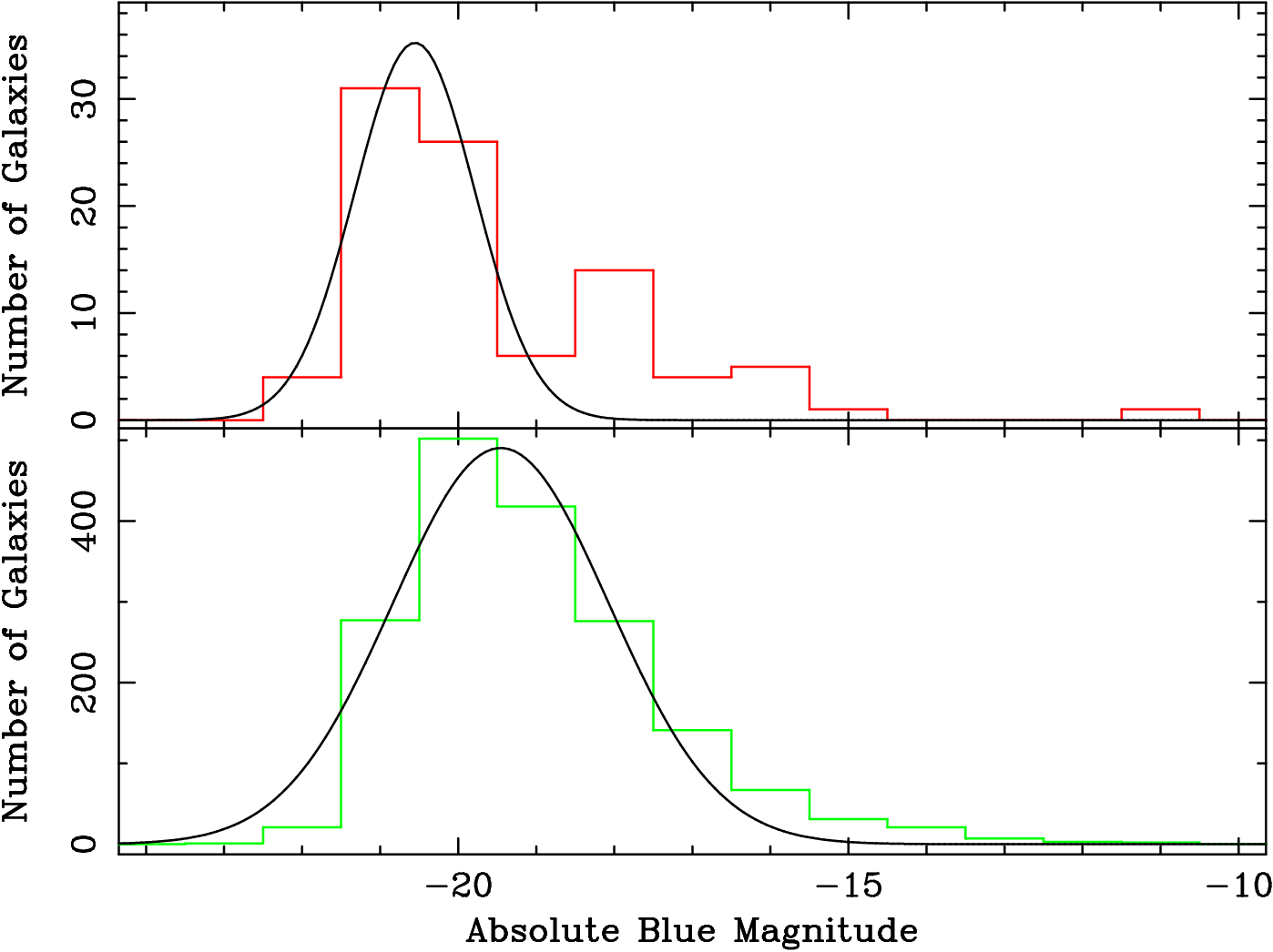}
\figcaption{Distribution of blue luminosities of the \cxo\ sample ({\em top})
and Tully catalogue of galaxies.}
\end{center}

Another means of displaying the dependence of the two samples on blue 
luminosity is shown in Figure~2. The mean absolute blue magnitude of the 
CXO sample is -20.6 or slightly brighter than 
that of the Tully catalogue at -19.4.

The CXO sample of galaxies spans the entire range of Hubble morphological 
types (Figure 3). There are 32 elliptical and lenticular galaxies and 50 
spiral and irregular galaxies in the sample.
%
\begin{center}
\includegraphics[angle=-90,width=5.0in]{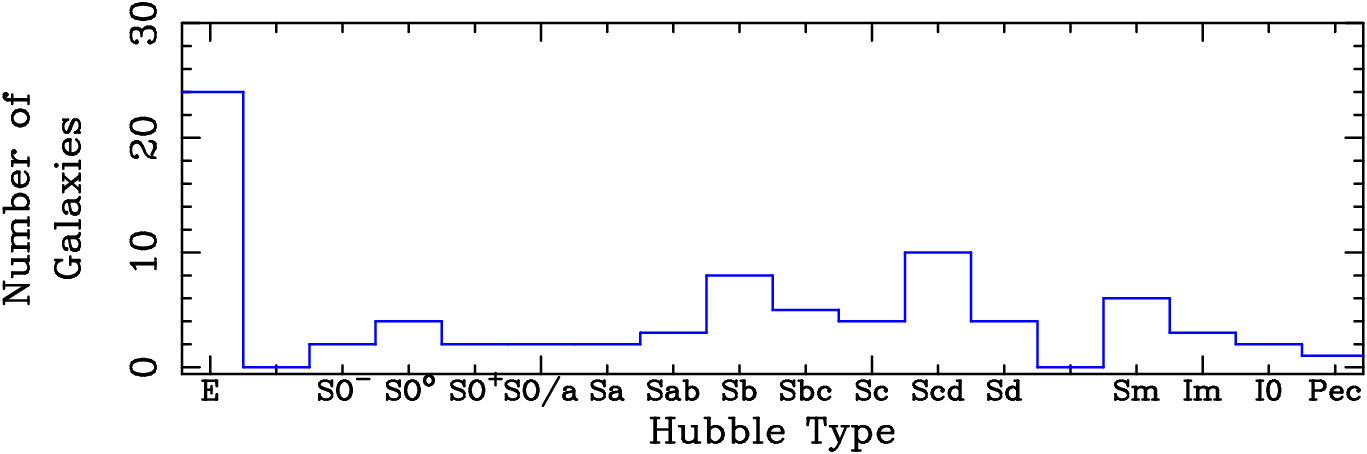}
\figcaption{Distribution of Hubble morpholigical types among
 the \cxo\ sample of galaxies.}
\end{center}

The distribution of the CXO sample of galaxies in blue and far-infrared (FIR) 
luminosity space is shown in Figure~4. The ellipticals are clustered near the 
upper 
left indicative of a relatively high mass per unit star-formation-rate (SFR) 
whereas the spiral galaxies generally span from lower-left to upper right in 
this plot going generally from small, late-type galaxies toward more massive 
early-type spirals. Note the several exceptions such as the starburst galaxy 
M82 (\lfir $=10^{44}$ , \lb $=2\times 10^{42}$).
%
\begin{center}
\includegraphics[angle=-90,width=5.0in]{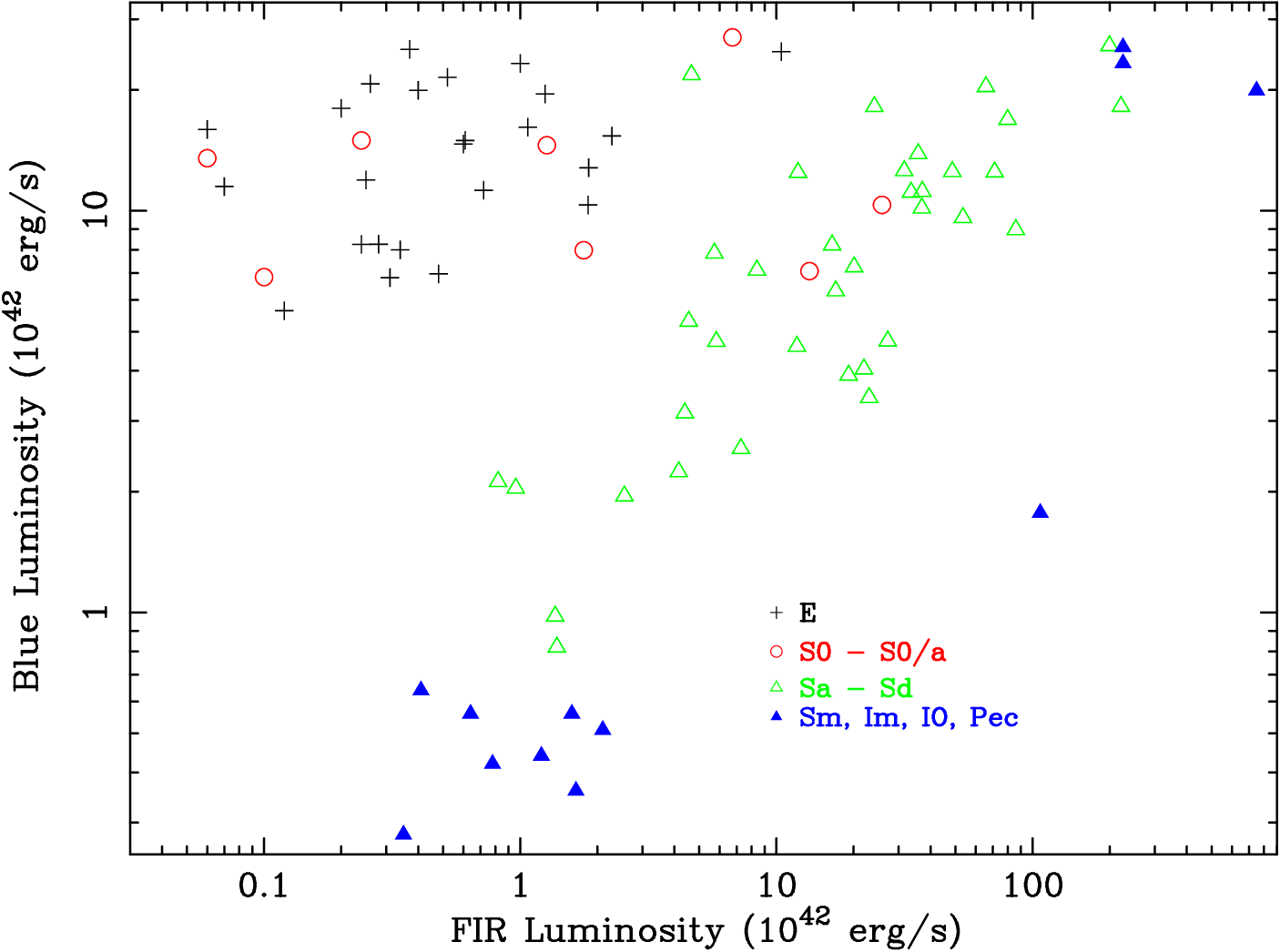}
\figcaption{Distribution of the \cxo\ sample of galaxies in $B$ and FIR 
luminosity.}
\end{center}

\section{Correlations with Galaxy Properties}

We have searched for correlations between the number and luminosities of ULX 
sources with numerous galaxy properties. Examples are shown in Figures~5 and 6. 
The distribution of X-ray luminosities of  individual ULXs 
against the host galaxy's FIR luminosity is shown above and against the B 
luminosity is shown below. There is a distinct separation between elliptical 
and spiral galaxies in FIR luminosity but that simply reflects the sample 
(see Figure~4). There is also a trend 
toward brighter ULXs in the spiral galaxies. However, there is also more ULX 
candidates in those galaxies and this trend may simply be a statistical 
fluctuation. Similarly, in blue light, there are more and brighter ULX 
candidates per galaxy in more massive (higher $B$ luminosity) galaxies. 
One expects to find more ULXs in more massive galaxies and perhaps, again, 
the trend toward more luminous ULXs is simply a statistical fluctuation. 

\begin{center}
\includegraphics[angle=-90,width=5.0in]{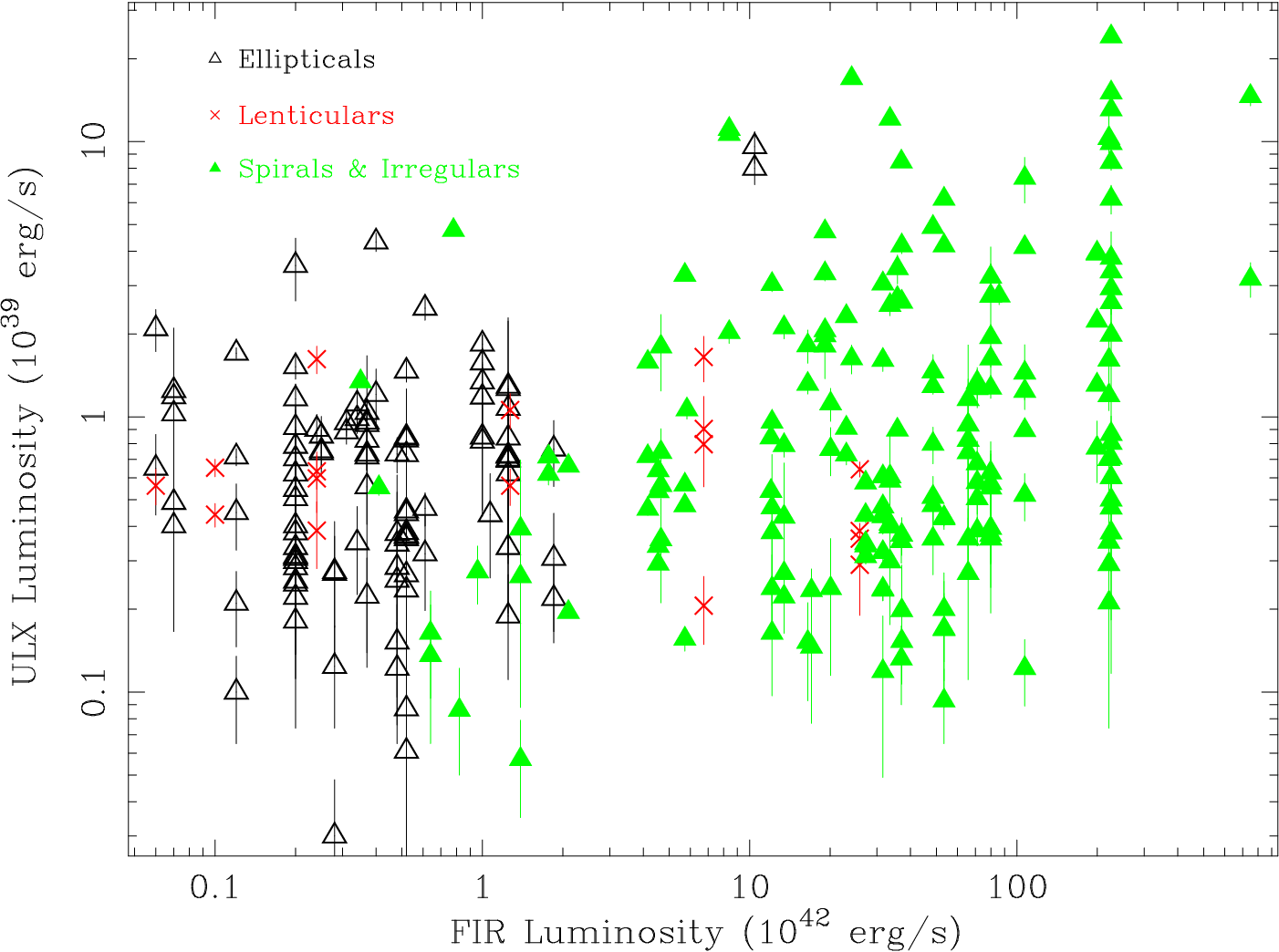}
\figcaption{Distribution of the ULX X-ray luminosities with host galaxy
FIR luminosity.}
\end{center}

\begin{center}
\includegraphics[angle=-90,width=5.0in]{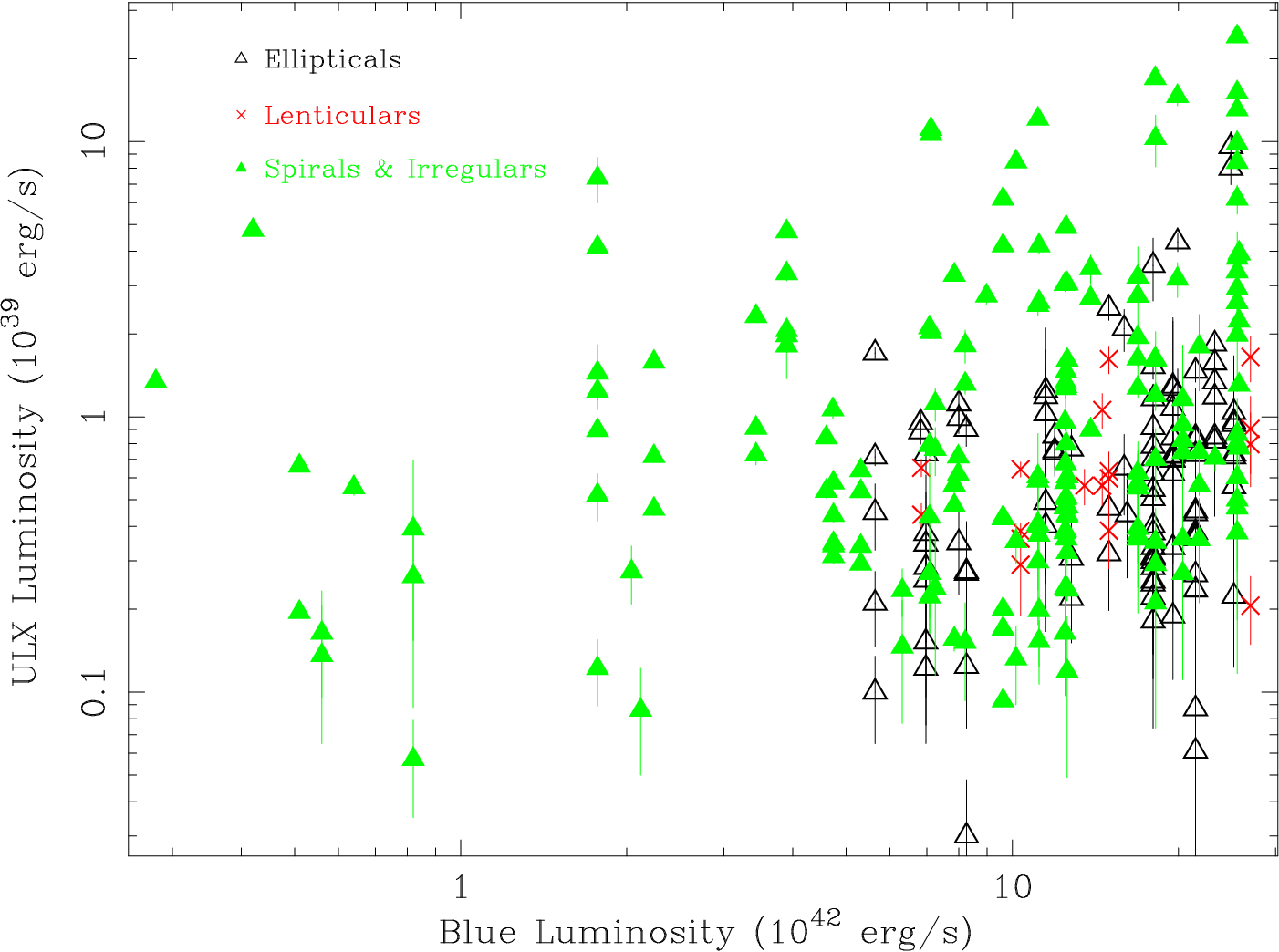}
\figcaption{Distribution of the ULX X-ray luminosities with host galaxy
$B$ luminosity.}
\end{center}

\begin{figure*}[h]
\begin{center}
\hspace{30pt}
\includegraphics[angle=-90,width=2.7in]{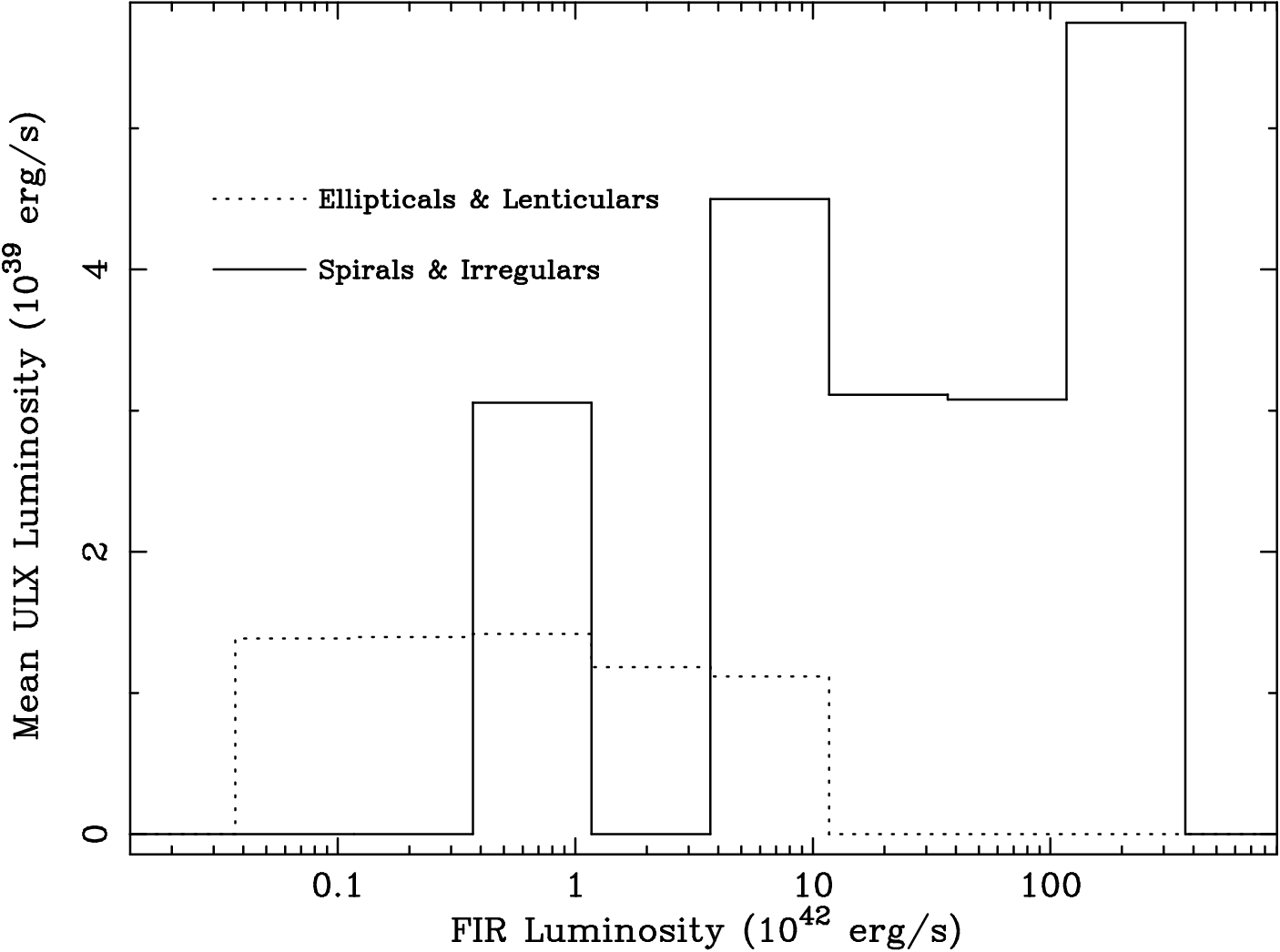}
\hspace{30pt}
\includegraphics[angle=-90,width=2.7in]{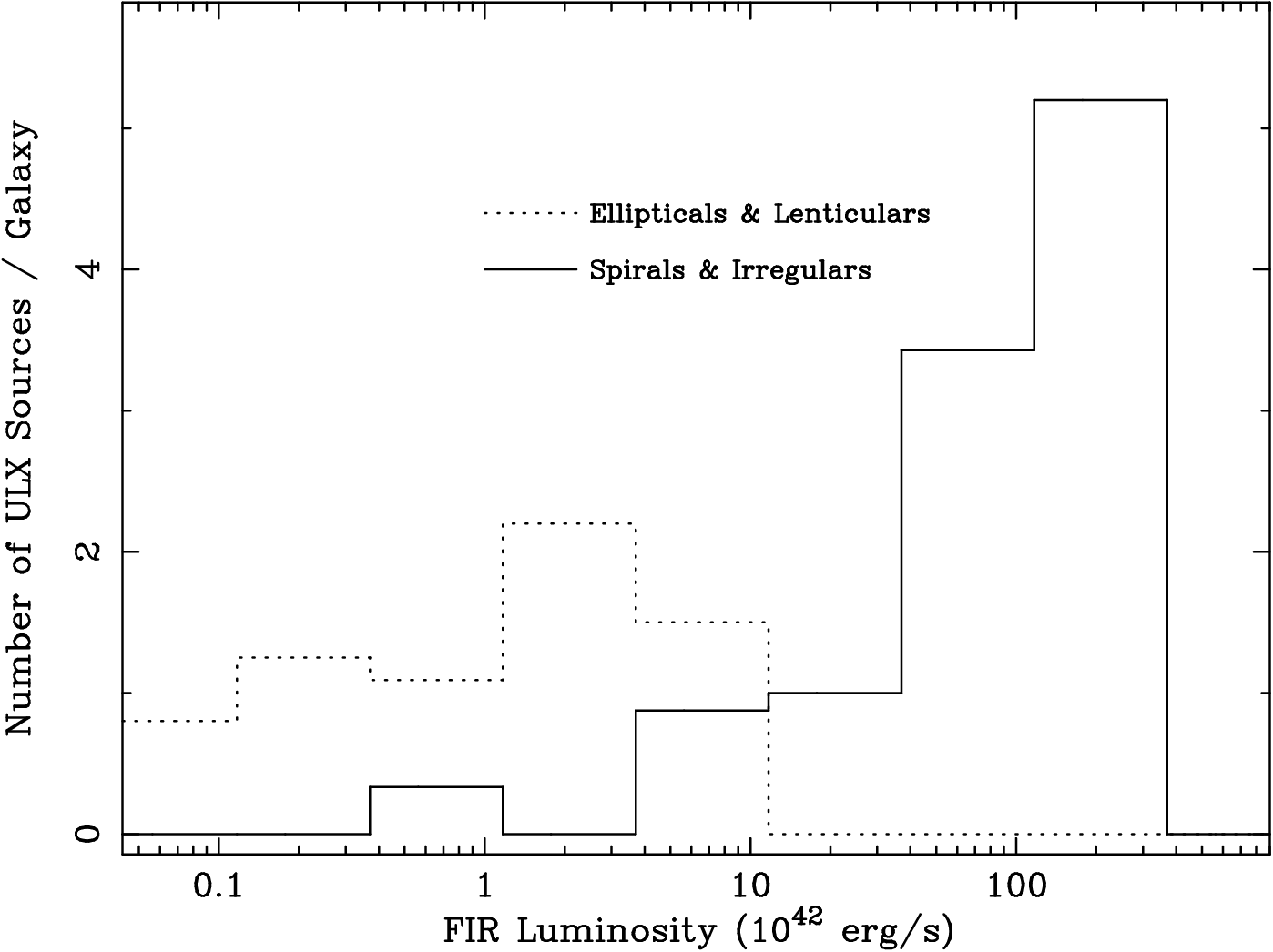}
\figcaption{Mean ULX luminosity ({\em left}) and the number of ULX candidates
per galaxy ({\em right}) against the FIR luminosity of the host galaxies.
}
\end{center}
\end{figure*}

The possible correlation with FIR luminosity is shown in a different way
in Figure~7. 
Here, only the ULX sources with \lx $>10^{39}$ erg/s are displayed. 
The figure at left shows that the brightest ULXs are indeed in the brightest 
FIR galaxies (although the distribution about the mean is broad, see  
Figure~6). The mean X-ray luminosity of ULXs in ellipticals is less than 
that in the spiral galaxies. At right we show that the number of ULXs per 
unit galaxy is strongly correlated with FIR luminosity. 
\begin{figure*}[h]
\begin{center}
\hspace{30pt}
\includegraphics[angle=-90,width=2.7in]{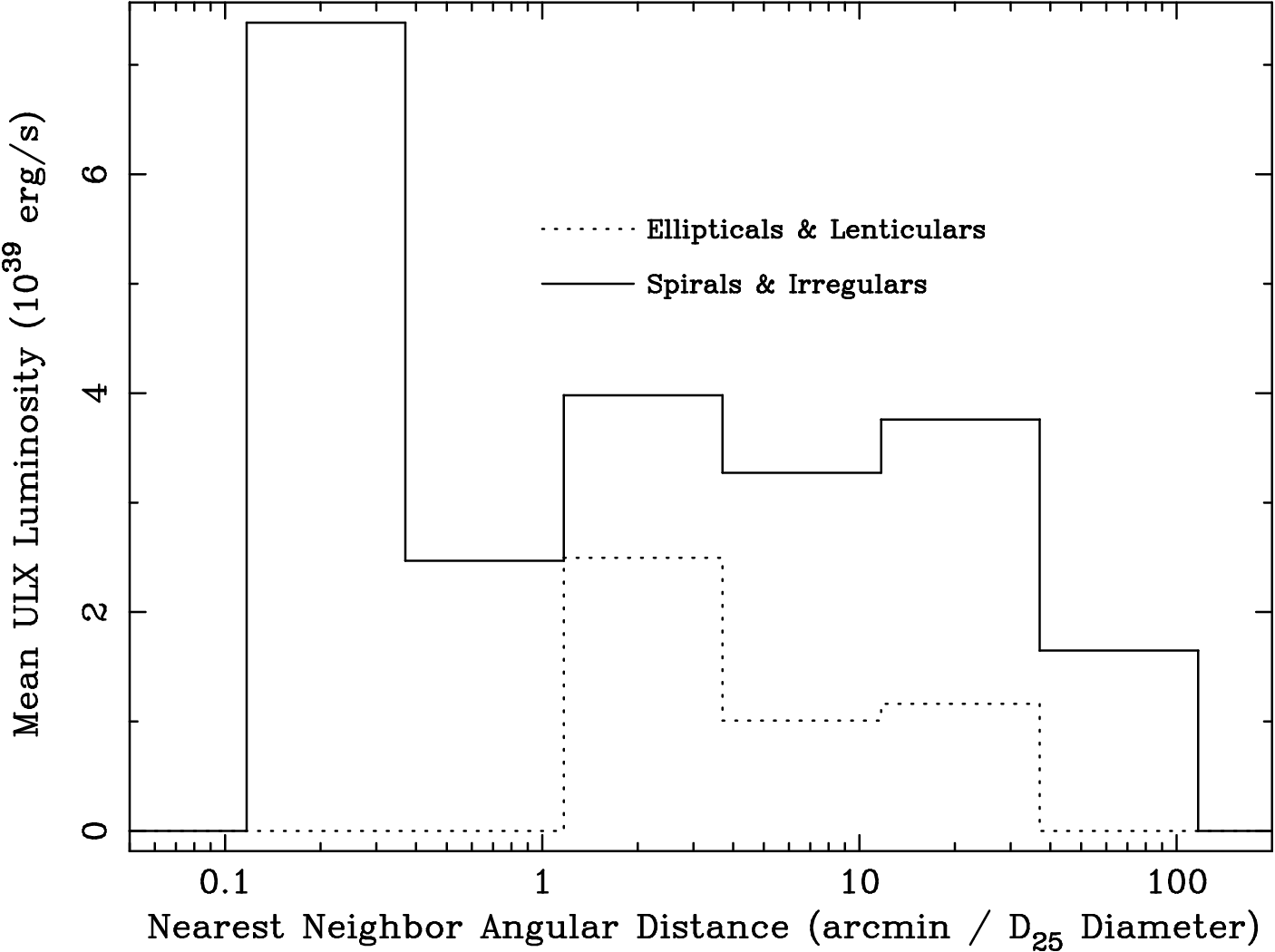}
\hspace{30pt}
\includegraphics[angle=-90,width=2.7in]{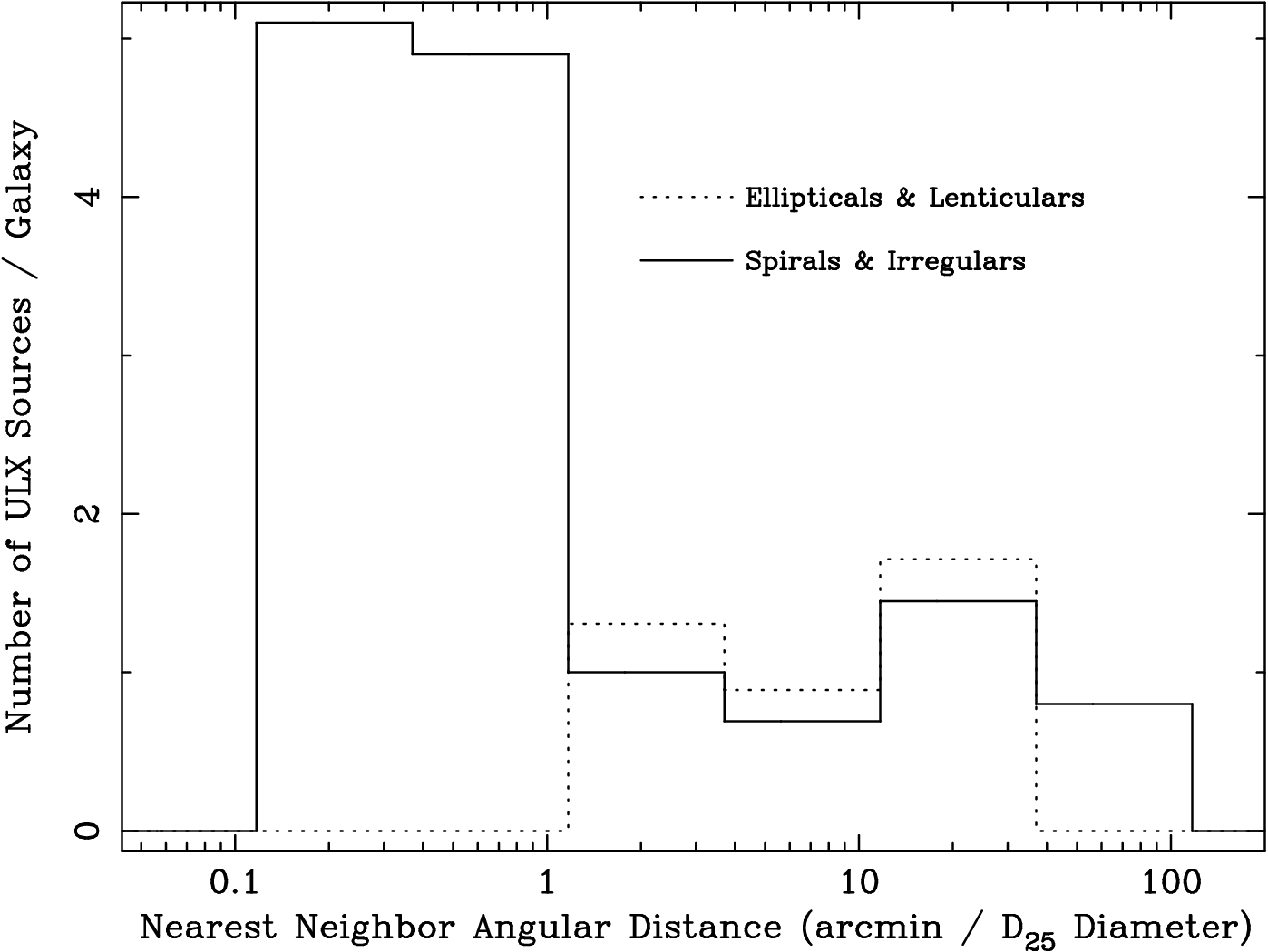}
\figcaption{Mean ULX luminosity ({\em left}) and the number of ULX candidates
per galaxy ({\em right}) against the distance to the 
nearest-neighbor of the host galaxies.}
\end{center}
\end{figure*}

In Figure~8 we show the only other strong correlation we have found between the 
number and luminosities of ULX sources with galaxy properties, namely, 
the distance to the host galaxy's nearest neighbor (measured in units of 
the \elip\ isophotal diameter). Those with seperations $<$1 are in the process 
of merging. They show much more activity (and high FIR luminosity, typically) 
and tend to have many ULXs. Ellipticals and spirals with more distant neighbors 
have much fewer ULXs and they are not as luminous although those in spirals are 
still more luminous than those in ellipticals.

\section{ULX X-Ray Spectral Properties}

\vspace{10pt}
\begin{center}
\includegraphics[angle=-90,width=5.0in]{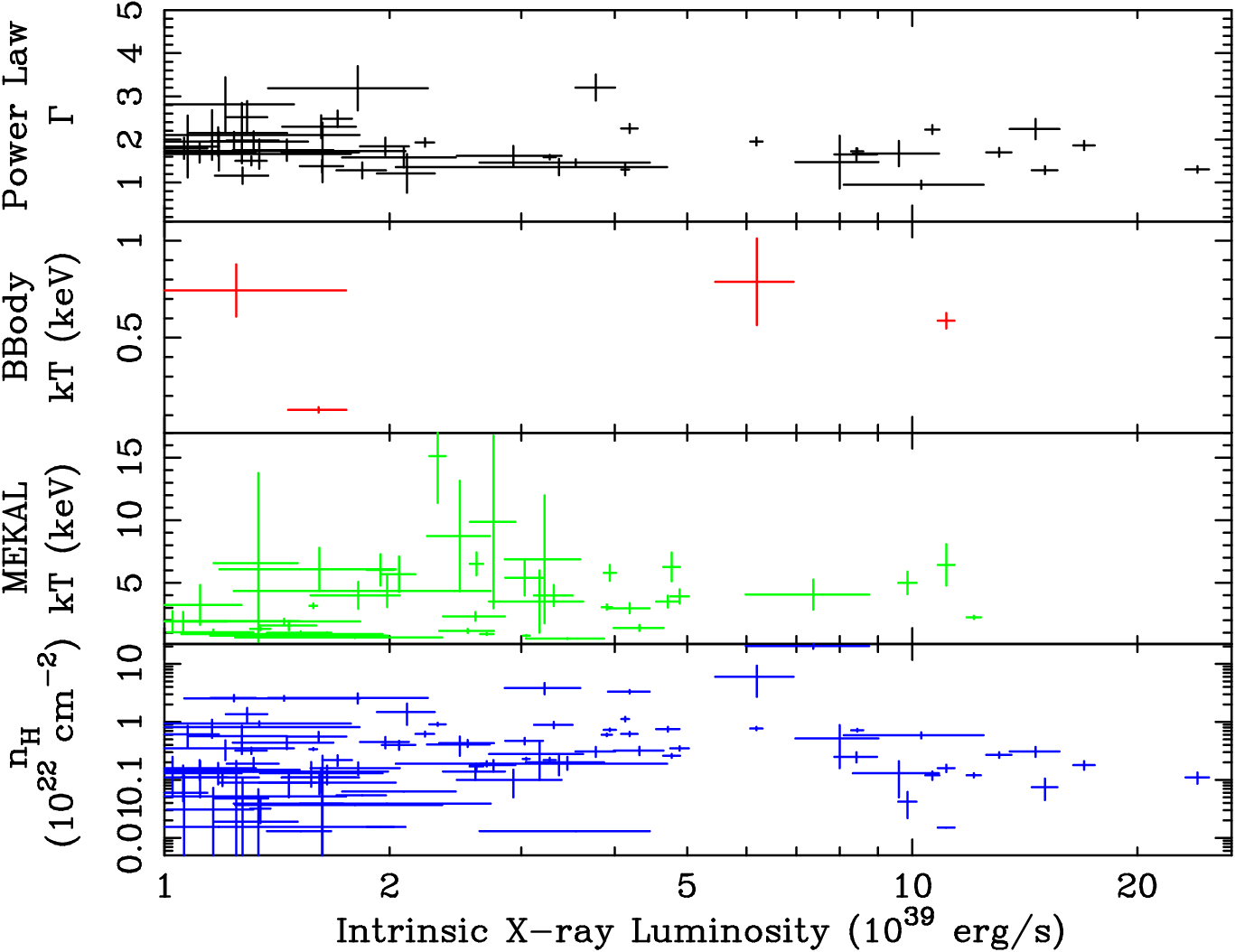}
\vspace{10pt}
\figcaption{Distribution of the ULX X-ray spectral parameters.}
\end{center}

In figure~9 is shown the distribution of the basic spectral parameters with the ULX 
source luminosities. Three simple models were used to characterize the 
0.5~--~8.0 keV spectra of the ULX candidates. Statistically, about 50\% were 
best-fit using an absorbed power law (mean spectral index for the group 
is 1.67) and about 50\% are absorbed thermal emission spectra 
(mean temperature 2.9 keV). Only 5 or about 4\% were best represented by an 
absorbed blackbody spectrum. Of course, many of the higher count sources 
will require more complex spectral models. This is work in progress. 
So far, no obvious correlation has been found between spectral properties 
and ULX luminosity nor with galaxy properties. 

\section{X-ray Luminosity Distributions}

\vspace{10pt}
\begin{center}
\includegraphics[angle=-90,width=5.0in]{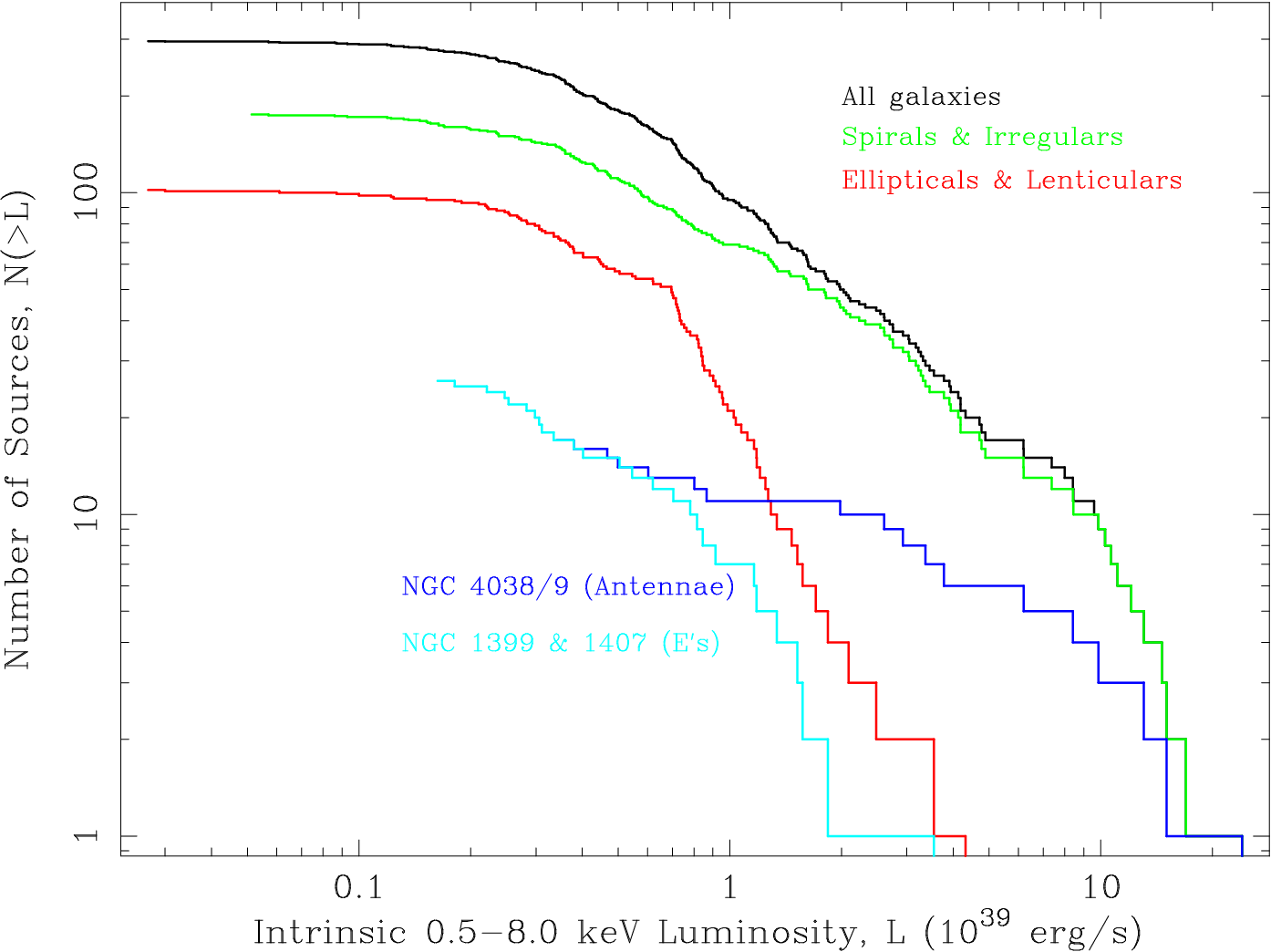}
\vspace{10pt}
\figcaption{LogN-logS distributions for the sample of galaxies.}
\end{center}

The luminosity functions of the entire CXO sample of ULX candidates is  
shown in Figure~10 along with the luminosity functions of the spirals and ellipticals. 
Also shown are a few of the galaxies with particularly high numbers of ULXs per 
galaxy. The horizontal scale has been extended to well below the ULX lower 
limit of $10^{39}$ erg/s to clearly show the incompleteness at low luminosities 
that results from our focus on the brightest objects. There are several 
features of interest. Clearly, the highest ULX luminosities in the spiral 
galaxies are much higher than those of the ellipticals and spirals dominate 
the total luminosity function throughout the entire range \lx $>10^{39}$. 
(However, only $\sim$3/8 of the galaxies are ellipticals and lenticulars, 
a factor that has not been accounted for here.) At the very highest 
end, the spirals are dominated by the single galaxy pair NGC 4038/9, 
with about 40\% of the ULXs brighter than $5\times10^{39}$ erg/s within 
that one system. The ellipticals exhibit a very different luminosity function. 
The slope of the elliptical galaxy luminosity function above $0.7\times10^{39}$ erg/s 
is -2.3 compared to -0.9 for spirals in the range $(0.7-10)\times10^{39}$ 
and to -1.0 
for the entire sample on that range. Above $10^{40}$ erg/s, there is a steep 
decline or cutoff in the luminosity function of the sample. There is also 
a distinct change in slope in the luminosity function of the elliptical 
galaxies at $0.7\times10^{39}$ erg/s. This is a real feature as opposed to the 
gradual flattening at low luminosities due to incompleteness (occuring in 
most of the curves at around $(2-4)\times 10^{38}$ erg/s). 
Because of this incompleteness, 
it is not possible to distinguish any change in slope at $2\times10^{38}$, 
the Eddington limit for typical neutron star accretors. 

\vspace{10pt}
\begin{center}
\includegraphics[angle=-90,width=5.0in]{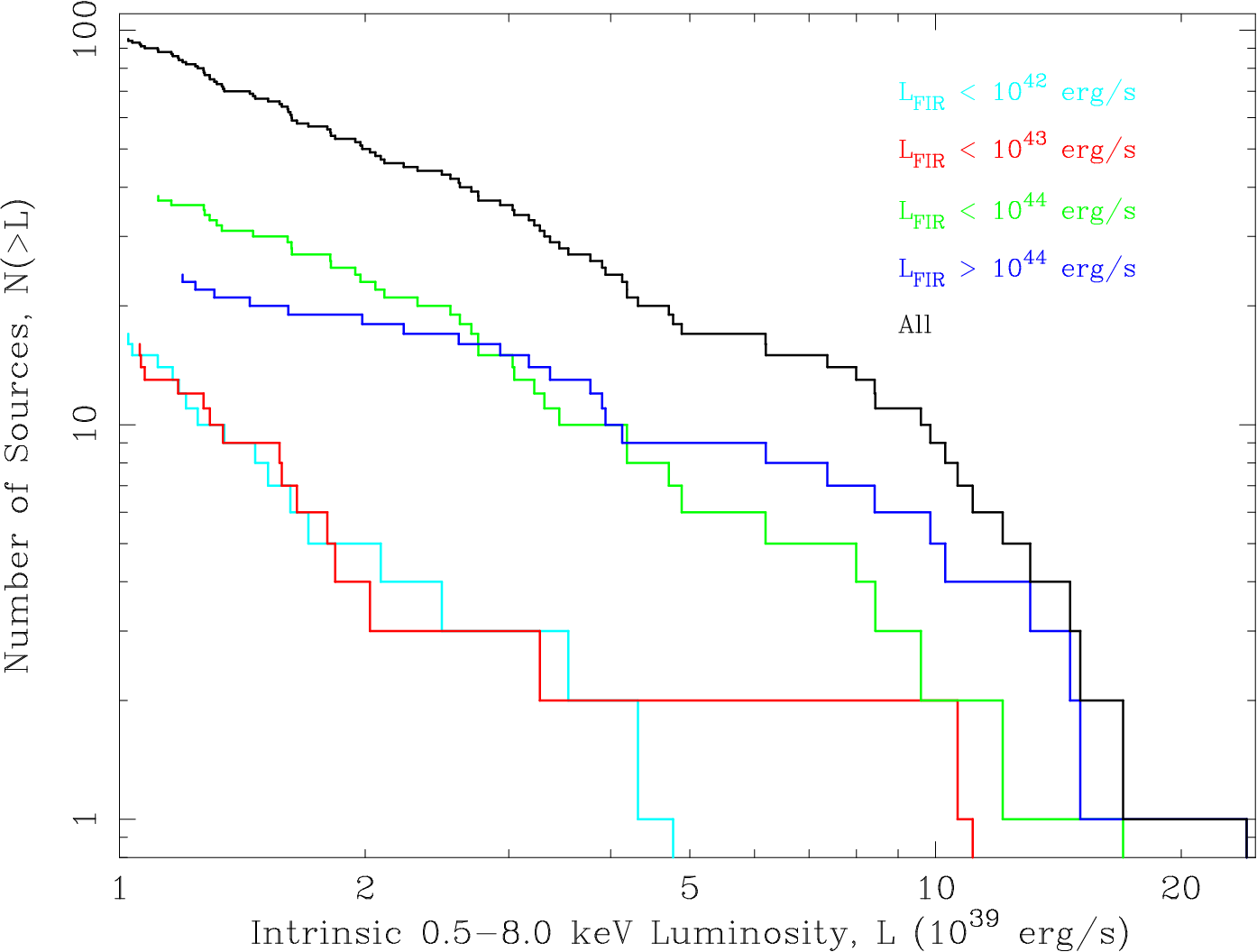}
\vspace{10pt}
\figcaption{LogN-logS distributions for subsamples of the galaxies with 
different FIR luminosities.}
\end{center}

The correlation of ULX X-ray luminosity with host galaxy FIR luminosity 
discussed previously is evident in the luminosity functions shown in
Figure~11. 
Here, the CXO sample has been partitioned into FIR luminosity ranges. 
With the exception of 2 very bright sources in the 
\lfir $= 10^{42}$ to $10^{43}$ erg/s range, there is a definite trend with the 
brightest ULXs occuring in the FIR-brightest hosts. (The 2 ``exceptions''
are both located in a galaxy with \lfir $= 8.4\times10^{42}$, 
the brightest in the group and just below the upper limit of the range.)

\section{Summary and Future Work}

We have analyzed a sample of 85 galaxies available in the Chandra archive 
of ACIS images. About 120 ULX candidates have been identified and their 
spectral properties determined. Few correlations have been found so far 
between the number and luminosities of the ULX candidates and properties 
of the host  galaxy population. Two correlations described here are with 
the galaxy's FIR luminosity and with the galaxy's interacting or merging 
status as measured by the distance to its nearest neighbor. 

Future research will include:
\begin{itemize}
\item Further searches for correlations with other galaxy properties including
the distributions of sources within the galaxy, correlations with spiral arms, 
bulges, bars, globular clusters, etc., and AGN activity.
\item Refined modeling of the observed X-ray spectra, particularly of the
highest count-rate sources where this is justified statistically.
\item Source position refinement by registering to accurately-known positions
of optical objects in the field, although Chandra positions are typically known to 
$\sim$0.6 arcsecs.
\item Extension to other wavelengths, particularly using archival HST and radio
data.
\item Examination of sample biases and completeness.
\end{itemize}

\end{document}